# Anomalous planar Hall effect in a kagome ferromagnet

Neeraj Kumar[1] and Y. Soh[1*],

1. Paul Scherrer Institut, 5232 Villigen, Switzerland

Yihao Wang[2], Junbo Li[2], and Y. Xiong[2]
2. Anhui Province Key Laboratory of Condensed Matter Physics at Extreme Conditions, High Magnetic Field Laboratory of the Chinese Academy of Sciences, Hefei 230031, China

**Abstract:**

The macroscopic signature for Weyl nodes so far has been the negative longitudinal magnetoresistance arising from the chiral anomaly. However, negative longitudinal magnetoresistance is not unique to chiral anomaly and can arise due to completely different mechanisms such as current jetting or in ferromagnetic systems due to the suppression of scattering with magnons. Therefore, a macroscopic effect that can be uniquely attributed to the presence of Weyl nodes is desirable. Here we show that the planar Hall effect could be a hallmark for Weyl nodes. We investigated the anisotropic magnetoresistance and planar Hall effect in $Fe_3Sn_2$, which has a kagome lattice and has been predicted to be a type II Weyl metal. We discover that the planar Hall effect contains a field antisymmetric contribution in addition to the ordinary field symmetric contribution. The field antisymmetric planar Hall effect has a 3-fold rotational symmetry, distinctively different from the symmetric planar Hall effect, but consistent with the 3-fold rotational degeneracy of the magnetization in $Fe_3Sn_2$. The temperature and field dependence of the antisymmetric planar Hall effect rules out an interpretation based on contribution from the anomalous Hall effect and is different from the symmetric planar Hall effect, pointing to a different origin. We attribute the antisymmetric planar Hall effect to the topological nature of $Fe_3Sn_2$ due to the presence of Weyl II nodes. Our finding offers a promising route for macroscopically probing Weyl systems.

**Introduction:**

There has been a large amount of excitement on topological materials due to the protected nature of the electronic band structure that can give rise to novel electronic properties. A macroscopic signature of topology in field induced quantum Hall systems[1], Dirac materials[2, 3], and two-dimensional topological insulators[4, 5] is the quantized Hall conductance, which can be uniquely identified due to the unusual quantization and exact expected values of the quantization. The analogous topological signature for Weyl semimetals has been the negative longitudinal magnetoresistance attributed to chiral anomaly[6]. However, in contrast to the quantized Hall conductance, negative magnetoresistance can arise from non-topological origins, such as due to current jetting[7] or in ferromagnets, due to the suppression of magnon scattering under the application of a magnetic field[8]. Therefore, it is not easy to distinguish the effect due to chiral anomaly from suppression of magnon scattering in the magnetoresistance of ferromagnets. Moreover, it has been predicted that in type II Weyl materials[9], the magnetoresistance due to the presence of Weyl nodes sometimes can be positive rather than negative. Therefore, there is a need for a new macroscopic hallmark for the presence of Weyl nodes, particularly in ferromagnets or in type II Weyl systems.

Recently, large anomalous Hall conductivity (AHC) and unusually large Hall angle has been suggested as supporting the presence of Weyl nodes in ferromagnets[10]. However, large anomalous Hall conductivity can be observed in ferromagnets and therefore these quantities themselves are not conclusive in determining the presence of Weyl nodes. In the case of the anomalous Hall effect (AHE), it needs to be examined cautiously to determine whether it is of extrinsic or intrinsic origin[11] as well as whether it is of topological origin[12], which itself can be due to real space spin texture[13] or reciprocal space Berry curvature. Therefore, determining the presence of Weyl nodes based on the AHE is not straightforward.

A natural effect to consider in ferromagnetic Weyl systems is the anisotropic magnetoresistance (AMR) and the planar Hall effect (PHE). The AMR is the resistance that depends on the direction of the magnetization with respect to the current. The PHE refers to the transverse voltage developed when an in-plane magnetic field is applied to the sample. It is worth comparing it against the ordinary Hall effect (OHE), which is the transverse voltage developed when an out-of-plane magnetic field is applied to the sample due to the carrier deflection arising from the Lorentz force and the AHE, which depends on the magnetization or Berry curvature. While the OHE is observed in most materials and its magnetic field dependence is used to extract information about the type of carriers and carrier density, the PHE is observed in systems with strong spin-orbit coupling.

The PHE was first detected in Ge[14], which is not ferromagnetic and subsequently in permalloy and Ni[15], in ferromagnets. It arises from the 4$^{th}$ term in the phenomenological equation describing the current in a crystal under the application of an electrostatic and magnetic field up to terms linear in $\vec{E}$ and quadratic in $H$:

$$\vec{j} = \sigma_0 \vec{E} + \alpha \vec{E} \times \vec{H} + \beta H^2 \vec{E} + \gamma (\vec{E}.\vec{H})\vec{H} + \delta \widetilde{M} \vec{E} \qquad (1)$$

, where $\vec{H} = (H_1, H_2, H_3)$ and $\widetilde{M} = \begin{pmatrix} H_1^2 & 0 & 0 \\ 0 & H_2^2 & 0 \\ 0 & 0 & H_3^2 \end{pmatrix}$.

The above equation was derived for a crystal with cubic symmetry. The symmetry of the crystal is relevant for the last term, which states that if $\delta \neq 0$, the current will depend on the orientation of the crystal. Since it is more convenient to fix the current and magnetic field in experiments and measure the voltage drop, it is more useful to invert the equation above and obtain an expression for the electric field[14]:

$$\vec{E} = \rho_0 (\vec{j} + A\vec{j} \times \vec{H} + BH^2 \vec{j} + C(\vec{j}.\vec{H})\vec{H} + D\widetilde{M}\vec{j}] \ . \qquad (2)$$

The 4$^{th}$ term gives rise to the AMR and PHE and shows that it is symmetric under magnetic field reversal and has a $\cos^2\theta$ and $\cos\theta\sin\theta$ dependence for the AMR and PHE, respectively, where $\theta$ is the angle between $\vec{j}$ and $\vec{H}$ and scales as $H^2$. In the case of ferromagnets, $\vec{H}$ is replaced by the magnetization $\vec{M}$. The AMR and PHE is mostly observed in ferromagnetic materials such as permalloy, Ni, Fe[16], GaMnAs[17], and $Fe_3Si$[18] and can be understood based on an anisotropic conductivity that depends on the magnetization direction owing to the spin-orbit coupling. Using a single magnetic impurity picture, a conducting electron scattering from the magnetic impurity will experience a different scattering cross section depending on the occupied orbital of the magnetic impurity, which depends on the magnetization direction due to the spin-orbit coupling. In most ferromagnetic materials, the scattering rate is higher when the magnetization is parallel to the electric current than when it is perpendicular. However, in some materials, including $Fe_3Sn_2$[19] and $Fe_3Si$, the opposite is true, i.e., the resistivity is lower in the direction parallel to the magnetization than perpendicular to the magnetization. Since the PHE is symmetric under the reversal of the magnetic field, it is not seen as a real Hall effect, which is antisymmetric under field reversal.

Recently, chiral anomaly in topological semimetals has been suggested as a contributing factor to the PHE[20, 21]. Topological materials such as $Cd_3As_2$[22], $ZrTe_5$[23], $NiTe_2$[24], $WTe_2$[25], and TaAs[26] are some of the examples where the PHE has been shown to exist. It has been proposed that in the case of type-II Weyl semimetals, an antisymmetric and linear in magnetic field contribution to the PHE can exist[9, 20]. However, the theoretical work on the PHE in Weyl systems did not consider systems with broken time reversal symmetry. It is worth noting that there has been no theoretical prediction of the magnetotransport properties associated with Weyl nodes in systems with broken time reversal symmetry due to the complexity of the problem.

We examine in detail the PHE in single crystal $Fe_3Sn_2$, a ferromagnetic metal composed of kagome bilayers which undergoes a spin reorientation transition (SRT) around 120 K[19, 27, 28] and has been proposed to have type-II Weyl nodes at the Fermi level[29]. We measured both the AMR and the PHE as a function of the azimuthal angle between the magnetic field and current, at various magnetic fields up to 9 T, and in the temperature range 2 – 300 K. Our key discovery is that in addition to the conventional field symmetric PHE, we observe an antisymmetric component, which we attribute to topological origin.

**Results and Discussion:**

Fig. S5(a) in the Supplementary Materials shows the schematic drawing of the measurement configuration along with the crystal structure of $Fe_3Sn_2$. The current is applied along the principal *a* axis in the kagome plane and $\theta$ is the angle of the magnetic field in the kagome plane with respect to the current. Resistance across each pair of longitudinal and transverse contacts is measured and used to calculate the AMR and the PHE. In the resistivity measured across the transverse contacts, several contributions were detected. First is the longitudinal resistivity term due to misalignment of the contacts. Second is the conventional field symmetric PHE ($\rho_{xy}^s$). Third is a contribution due to the difference between the sample plane and the field rotation plane, which contributes a Hall voltage proportional to the out-of-plane magnetic field. Last is a field antisymmetric PHE ($\rho_{xy}^a$). The details of extracting various quantities from the raw data is discussed in the Supplementary Materials.

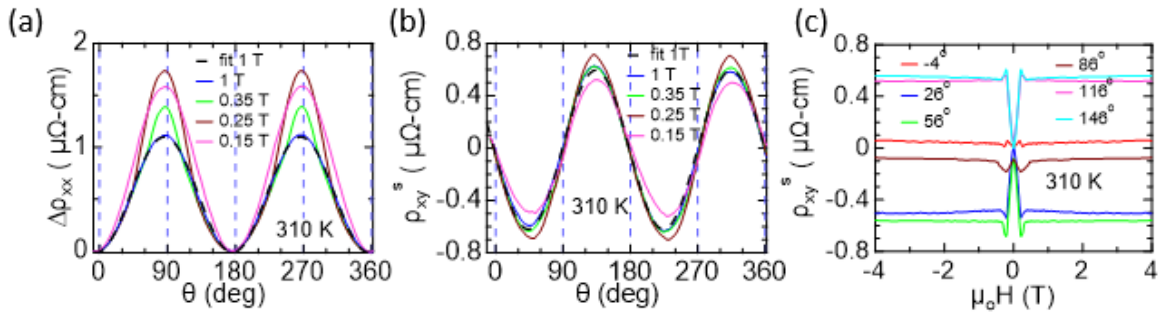

Fig 1. (a) Change in the longitudinal resistivity on rotation of magnetic field vector in the *ab*-plane at selected field magnitude. (b) corresponding planar Hall effect. (c) Planar Hall effect as a function of magnetic field at select angles.

Fig. 1 shows the transport data at 310 K. According to the conventional description of the PHE[14, 15]

$$\rho_{xy}^S = -\frac{1}{2}\rho_{xy}^S(0)sin2\theta_M \qquad (3)$$

$$\Delta\rho_{xx} = \rho_{xx} - \rho_\parallel = \Delta\rho_{xx}(0)\sin^2\theta_M \qquad (4)$$

, where $\theta_M$ is the angle of the magnetization $\vec{M}$ with respect to the current $\vec{j}$, $\rho_{xy}^S(0)$ is the peak to peak amplitude of the PHE $\rho_{xy}^S$ and $\Delta\rho_{xx}(0)$ is the peak to peak amplitude of the AMR $\Delta\rho_{xx}$. Fig. 1(a) shows the change in the longitudinal resistivity $\Delta\rho_{xx}$ vs $\theta$ at room temperature and various magnetic field values. Above saturation, $\theta_M = \theta$, thus $\Delta\rho_{xx}$ follows the standard $sin^2\theta$ variation on the expected lines as demonstrated by the fitting curve at 1 T. Below saturation, a small deviation is observed due to the difference between $\theta$ and $\theta_M$ caused by magnetic anisotropy. Furthermore, below the saturation field, $\Delta\rho_{xx}(0)$ scales as $M^2$ up to its maxima (see Fig. S6(a) in Supplementary Materials), where $M$ is the magnetization for $H \parallel \mathbf{a}$ [15]. Above saturation, $\Delta\rho_{xx}(0)$ is independent of the magnetic field strength as the system becomes fully polarized, as shown in Ref. [19]. $\Delta\rho_{xx}(0)$ is 1.14 μΩ-cm at 310 K and 9 T.

We first discuss the field symmetric PHE. Fig. 1(b) shows the symmetric component $\rho_{xy}^S$ of the PHE at several magnetic field values at 310 K, which follows the expected behavior $sin2\theta$ with the peak-to-peak amplitude $\rho_{xy}^S(0)$ of the PHE of 1.3 μΩ-cm. $\rho_{xy}^S(0)$ is close to $\Delta\rho_{xx}(0)$. It was shown in single crystalline Ge[14] and La$_{0.7}$Ca$_{0.3}$MnO$_3$ and La$_{0.75}$Sr$_{0.25}$MnO$_3$ films that the absolute magnitude of $\Delta\rho_{xx}$ and $\rho_{xy}^S$ and their ratio with respect to each other is affected by the direction of the electrical current with respect to the crystal axes, with extreme cases where the amplitude can vanish.[30, 31] As seen in Fig. 1(c), above the saturation field, the PHE amplitude is nearly independent of the magnetic field, similar to the case for the longitudinal resistivity. As with $\Delta\rho_{xx}(0)$, $\rho_{xy}^S(0)$ also scales as $M^2$ up to its maxima (see Fig. S6(b) in Supplementary Materials)[15].

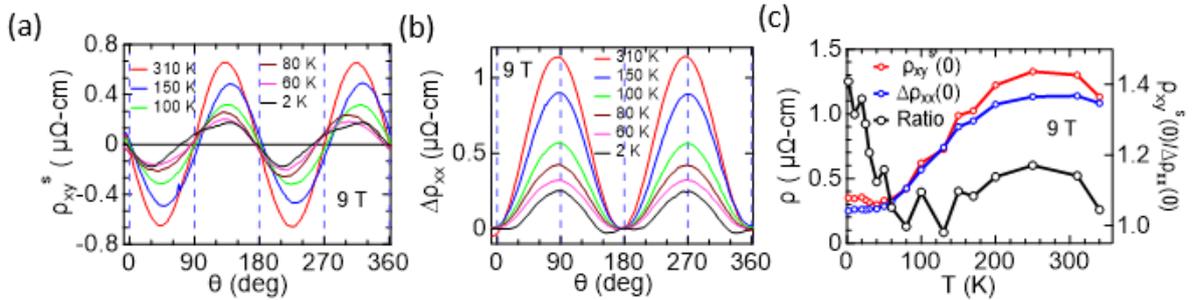

Fig 2. (a) Planar Hall effect curves at various temperatures at 9 T. (b) corresponding change in the longitudinal resistivity. (c) Variation in the peak to peak PHE and AMR at 9 T as a function of temperature, with the ratio of the PHE with AMR on right side scale.

We obtain the field symmetric component of the PHE and AMR at different temperatures at 9 T as shown in Fig. 2(a)-(b). The amplitude of the PHE decreases with decreasing temperature. Similarly, for AMR, the amplitude decreases with decreasing temperature as seen in Fig. 2(b). As summarized in Fig. 2(c), both quantities show overall a similar behavior. For both longitudinal and transverse channels, we observe a decrease in the amplitude with decreasing temperature until they level off at around 50 K, which is due to $\rho_0$ decreasing rapidly as the system is cooled down, since Fe$_3$Sn$_2$ is a metal with a very high residual resistivity ratio (defined as $\frac{\rho(300K)}{\rho(2K)}$) of 40. However, when we consider instead the normalized amplitude with respect to $\rho_0$, which is the term $C(\vec{J}\cdot\vec{H})\vec{H}$ in Eq. 2, it increases by 13 and 9 fold for the PHE and AMR, respectively, as we lower the temperature, since $C$ depends on the carrier lifetime, which increases as the system is cooled down. The normalized amplitudes obtained at different temperatures plotted as a function of $\sigma_o$ instead of temperature, shows a monotonic increase, as expected, due to the increase of carrier lifetime (see Fig. S7(a) in Supplementary Materials). Their ratio showed in the inset of Fig. 2(c) is nearly constant from 300 to 100 K and increases below 100 K. The change in the amplitudes as well as the ratio of the amplitude of the AMR to the amplitude of the PHE as the system undergoes a SRT around 120 K could reflect a fundamental change in the electronic structure of $Fe_3Sn_2$ at the Fermi level below the SRT compared to that above the SRT[28]. This could be due to a shift of the chemical potential, which then causes different electronic states to intercept the chemical potential. However, it is also possible that the increase in the ratio of $\frac{\rho_{xy}^S(0)}{\Delta\rho_{xx}(0)}$ is due to extra PHE caused by the Weyl points[21]. A similar increase in the PHE is seen also in GdPtBi, another Weyl semimetal[32].

While the field symmetric component of the PHE in Fig. 2(a) continues to show a 2-fold symmetry, we observe a change in the shape of the curve at low temperatures especially below 60 K. The angle at which the PHE shows minima decreases with decreasing temperature (see Fig. S5 (b)). The PHE at 2 K and at 9 T (see Fig. 3(a)) shows almost a saw tooth shape. This is clearly different from the sinusoidal shape seen at room temperature. The curve clearly shows inflection points at 30° and at 150°, which are high symmetry crystallographic directions in the kagome plane, suggesting that the PHE is dependent on the in-plane crystal structure. Furthermore, the PHE curve crosses zero at 72° instead of at 90° such as at room temperature. The PHE, however, shows a change in slope at 90°. The shape of the PHE curve is strikingly similar to the data seen for $YRh_6Ge_4$, where it is attributed to the involvement of longitudinal magnetoresistivity[33]. In our case, we have, however, carefully removed any longitudinal contributions. A similar saw-tooth shape of the PHE seen in TaP was attributed to current jetting in a high mobility system[34]. However, in $Fe_3Sn_2$, the mobility is not very high, and thus the saw tooth shape cannot be ascribed to the current jetting either. In our case, the shape of the AMR curves also changes below 60 K deviating from the sinusoidal form and the minima splitting into two. This could mean a change in the electronic structure of $Fe_3Sn_2$, but the change in the shape should take into account the larger mean free time at lower temperatures, which enables a better probe of the Fermi surface as $\omega_c\tau\ (=\mu H)$ increases, which is masked at higher temperature due to the shorter mean free time. Especially, these effects are accessed at low temperature and at high field, as shown in Fig. 3(a) and (b). A similar change in the shape is also seen in manganites.[30]

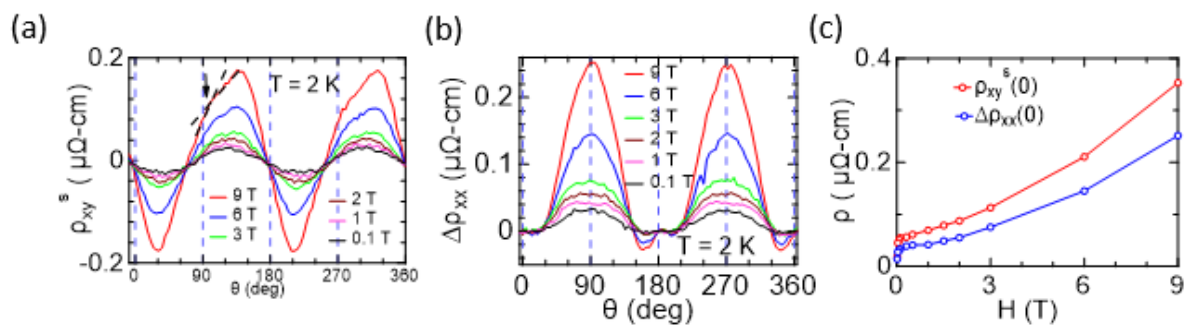

Fig 3. (a) Planar Hall effect curves at 2 K for various magnetic field strengths. (b) The AMR curves at the corresponding fields. (c) Net PHE and AMR as a function of the field.

Now we look at the evolution of the angular dependence of the field symmetric component of the PHE and AMR depending on the magnetic field at low temperature. Fig. 3(a) shows the angular dependence of the PHE at 2 K at various magnetic field values. The PHE increases gradually and the shape deviates from sinusoidal at high fields. The dependence of the PHE with field at 2 K is much stronger compared to the corresponding dependence at 300 K (see Fig. 1(c) and Fig. S8(b) in Supplementary Materials). The corresponding AMR is shown in Fig. 3(b). The AMR amplitude increases with field as well. The amplitude of the PHE and AMR are compiled in Fig. 3(c), with both showing similar power law dependence of $H^{1.6}$. A similar growth with exponent of 1.5 for the PHE and AMR was seen in the case of NiTe$_2$ [24], indicating that although the two quantities are not equal, they are proportional to each other. It is worth noting that the field and temperature dependence of the field symmetric PHE and AMR are similar suggesting that they arise from the same physical origin. Their ratio, however, changes with the field (see Fig. S5(c) in Suplementary Materials).

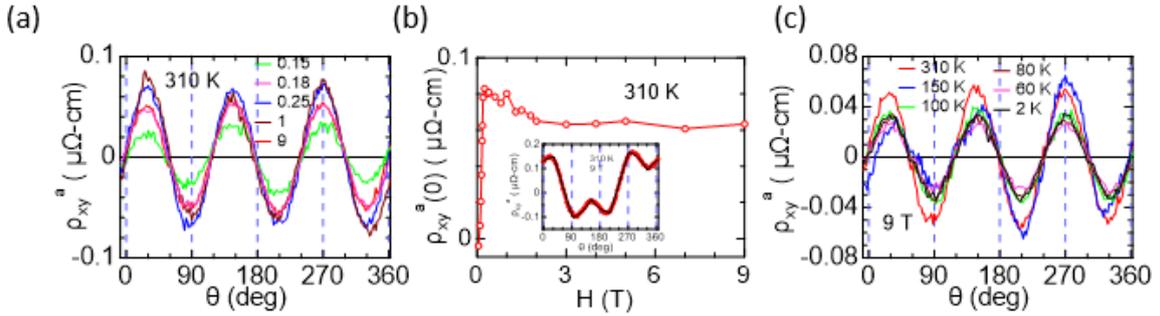

Fig 4. Antisymmetric planar Hall effect (a) at 310 K at selected field values. (b) Amplitude as a function of field. Inset: Total antisymmetric term across Hall contacts. (c) Antisymmetric planar Hall effect at 9 T at various temperatures.

Now we focus our attention on the field antisymmetric part $\rho_{xy}^a$ of the PHE. The inset in Fig. 4(b) shows the antisymmetric term across the Hall contacts before we factor out any antisymmetric contribution from out-of-plane Hall effect. The presence of an out-of-plane magnetization component is possible due to a small misalignment between the sample plane and field rotation plane, which should be sinusoidal in nature with 2π periodicity. For the Hall effect arising from an out-of-plane field or magnetization component, an AHE would mostly dominate the contribution in Fe$_3$Sn$_2$. A function consisting of an out-of-plane sinusoidal AHE term and an in-plane triangular term shows a good fitting to our antisymmetric Hall resistance. Instead of a triangular term, a $sin3\theta$ term would have fitted equally well. Both functions give qualitatively the same behavior, except that the value of $\rho_{xy}^a$ obtained from a triangular function fitting is 1.2 times that of the corresponding value obtained using a $sin3\theta$ function fitting. Supplementary Materials contains details of the fitting procedure. Based on our analysis, the tilt between the sample plane and field rotation plane is about 1º.

Fig. 4(a) shows the angular dependence of the antisymmetric part of the PHE $\rho_{xy}^a$ at 310 K and selected magnetic field values after factoring out the out-of-plane contribution. $\rho_{xy}^a$ at 310 K exhibits a 3-fold symmetry. We suspect that the 3-fold symmetry in $\rho_{xy}^a$ reflects the symmetry of the in-plane kagome crystal structure. The change in amplitude of the antisymmetric PHE $\rho_{xy}^a(0)$ with field is shown in Fig. 4(b). $\rho_{xy}^a(0)$ first increases with the field up to 0.25 T and decreases afterwards before leveling off above 3 T. At room temperature, up to its maximum value, $\rho_{xy}^a$ varies as $M^4$ (see Fig. S6(c) in Supplementary Materials), as opposed to the $M^2$ variation seen for $\rho_{xy}^s$. Fig. 4(c) shows $\rho_{xy}^a$ at 9 T and various temperatures showing the change in this term with temperature. This behavior is summarized in Fig. 5(c) at 1 and 9 T.

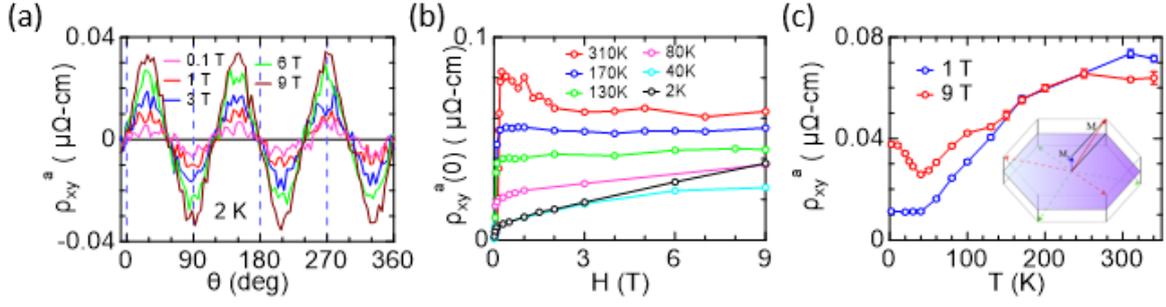

Fig 5. Antisymmetric planar Hall effect (a) at 2 K at various magnetic fields. (b) Field dependence of its amplitude at several temperatures. (c) Corresponding temperature dependence at 1 and 9 T.

At 2 K, $\rho_{xy}^a$ increases with magnetic field as seen in Fig. 5(a). The field dependence of $\rho_{xy}^a(0)$ is shown in Fig. 5(b) at 2 K and several other temperatures. At all temperatures, a spontaneous jump near zero field is observed and the value of the jump continues to decrease with decreasing temperature. However, if we plot instead the normalized value $\frac{\rho_{xy}^a(0)}{\rho_o}$ as a function of $\frac{H}{\rho_o}$, which is proportional to $\omega_c \tau$, the value of the jump is similar at all temperatures (see Fig. S7(c) in Supplementary Materials). At fields above which the initial jump occurs, the behavior gradually changes with decreasing temperature. At 310 K, between 0.25 and 3 T, there is a gradual decay of $\rho_{xy}^a(0)$ before it levels off at fields greater than 3 T. We observe this feature down to 230 K (extra data shown in Fig. S4 in Supplementary Materials). This decay at low fields is not seen at 170 or 130 K. At 80 K, we start to see an increase in $\rho_{xy}^a(0)$ with field. This slope of $\rho_{xy}^a(0)$ vs $H$ increases further at 2 K. This behavior is emphasized better in Fig. 5(c). At 1 T, the antisymmetric PHE decreases with temperature and levels off at 40 K. The curve at 9 T shows a similar behavior down to 150 K and starts to deviate at 130 K, signaling the appearance of a linear increase in $\rho_{xy}^a(0)$ with field. This incidentally coincides with the temperature at which the spin reorientation transition occurs in $Fe_3Sn_2$. At 40 K, the curve shows a minimum before increasing with decreasing temperature, which suggests an electronic transition. This temperature coincides with a transition seen in the AMR for $H \parallel \mathbf{c}$[19]. It is worth noting that if we normalize $\rho_{xy}^a(0)$ with respect to $\rho_0$ as we did for the symmetric PHE, the normalized amplitude, which is better representative of $\gamma$ in Eq. (1), increases by 23 fold as the system is cooled down. The normalized amplitude obtained at different temperatures plotted as a function of $\sigma_o$ shows a monotonic increase as in the case of $\frac{\Delta \rho_{xx}(0)}{\rho_o}$ and $\frac{\rho_{xy}^s(0)}{\rho_o}$ (Fig. S7(a)-(b)).

In order to understand the origin of our antisymmetric PHE, it is important to compare our $\rho_{xy}^a$ with the antisymmetric term in the PHE observed in $Fe_3Si$[18], GaMnAs[35], and Fe[36, 37] films. In all these cases, the films were grown on either vicinal surfaces or high index planes. In ref[35, 36], it is clear that the magnetic easy axis is not confined to the film plane and thus the antisymmetric Hall term (with $2\pi$ periodicity) could be due to the out-of-plane magnetization component. This kind of contribution is similar to that from the tilt of the sample with respect to the field rotation plane, which we have already subtracted. In Fe films[37], a term with only $2\pi$ periodicity is present and we believe it is due to an out-of-plane Hall effect, similar to ref. [36]. In ref. [18], the antisymmetric PHE term is only detected in low symmetry planes but not on $Fe_3Si$ (001). They assume that the magnetization lies in the plane with 4-fold symmetry[38] and rule out any contribution from out-of-plane magnetization. Their data on $Fe_3Sn$(113) thin films is fitted with a $C_1 cos\theta + C_2 cos^3\theta$ term and explained based on a phenomenological model that incorporates terms higher than $H^2$ in the second rank magnetoresistivity tensor. It is worth noting that we can fit their data with a 3-fold antisymmetric term such as ours after we subtract a sinusoidal part with $2\pi$ periodicity, even though their in-plane magnetic symmetry is 4-fold[38] in contrast to ours, which is 3-fold. An antisymmetric PHE derived from 3$^{rd}$ order direction cosines yields an expression where the coefficients obey $\frac{C_1}{C_2} \approx -1.2$. However, we find some limitations in their arguments. First, it is not very clear whether the $cos\theta$ term is of in-plane origin since a direct comparison of this term with an out-of-plane Hall effect is missing. Second, their fit of the data at 300 K gives a ratio of $\frac{C_1}{C_2} \approx -1.6$, which is close to their phenomenological model, but their fit at 77 K gives $\frac{C_1}{C_2} \approx 2.2$, which is of opposite sign. Third, such an expansion to higher order terms yields a higher order term also in the AHE, but they cannot detect it experimentally. Fourth, a phenomenological expansion to higher order term yields an antisymmetric term even for low index planes Fe$_3$Si(001), but they do not observe it. An anomalous like Hall effect is also reported in the case of Mn$_3$Sn, a non-collinear antiferromagnet, where the cause is proposed to be a small magnetic moment rotating around the magnetic field[39].

In Fe$_3$Sn$_2$, we are able to detect this 3-fold antisymmetric term in a high symmetry kagome plane. At first instance, we could consider this antisymmetric Hall effect arising from the fact that the in-plane easy axis is not strictly in the kagome plane but rather at a small angle around 30º[40] while still being 3-fold symmetric as shown in Fig. 5(c) inset. If we assume that the anisotropy energy for rotating the magnetization to the kagome plane is very high so that even at 9 T the magnetization is not entirely in the kagome plane, an azimuthal rotation of the field will cause the magnetization to switch from one easy axis slightly tilted off the kagome plane to another with 120º periodicity. However, this picture is not supported by the temperature dependence of the antisymmetric PHE. The AHE for $H \parallel$ **c** as a function of temperature decreases much more rapidly compared to the dependence seen in Fig. 5(c) (see Fig. S3(a)-(b) in Supplementary Materials for the AHE data). Furthermore, such a contribution from an out-of-plane magnetization component would decrease as we increase the strength of the in-plane field. While our data above 200 K shows such a trend, our data at temperatures below 130 K show an increase of the antisymmetric Hall effect as we increase the strength of the magnetic field, contradicting this picture. Thus, this in plane antisymmetric PHE cannot be explained by conventional models and points to a different origin.

We propose that the antisymmetric PHE that we discovered is the realization of an antisymmetric and linear PHE due to the presence of Weyl nodes near the Fermi level in a time reversal symmetry broken system. Based on DFT calculations on Fe$_3$Sn$_2$, a large number of Weyl nodes are predicted to exist near the Fermi level[29]. Recent theory on non-magnetic type II Weyl semimetals predict a $2\pi$ periodic sinusoidal angular dependence and linear PHE[9]. On the other hand, our system is a ferromagnet besides having type II Weyl nodes and therefore the PHE response should reflect contributions from both aspects. It is worth noting that the symmetry, field and temperature dependence of the antisymmetric PHE is distinctively different from that of the symmetric PHE or AMR pointing to a different origin. We believe that our field symmetric PHE contribution arises from Fe$_3$Sn$_2$ being a ferromagnet, whereas the field antisymmetric PHE contribution arises from the presence of type II Weyl nodes in the system. Since the magnetization has a 3-fold rotational degeneracy, the associated band structure and Weyl nodes determined by the magnetization direction owing to the spin-orbit coupling[29] will have accordingly a 3-fold rotational degeneracy giving rise to the 3-fold triangle antisymmetric PHE that we discovered.

**Methods:**

Details of the sample preparation and measurements are described in our previous paper[19].

**Acknowledgements:**

This project has received funding from the European Union's Horizon 2020 research and innovation programme under the Marie Skłodowska-Curie grant agreement No 701647. X. Yiong acknowledges funding support from the Natural Science Foundation of China, Grants No. U1432138; a portion of this work was supported by the High Magnetic Field Laboratory of Anhui Province.

# Anomalous planar Hall effect in a kagome ferromagnet

Neeraj Kumar[1] and Y. Soh[1*],

1. Paul Scherrer Institut, 5232 Villigen, Switzerland

Yihao Wang[2], Junbo Li[2], and Y. Xiong[2]
2. Anhui Province Key Laboratory of Condensed Matter Physics at Extreme Conditions, High Magnetic Field Laboratory of the Chinese Academy of Sciences, Hefei 230031, China

**Supplementary Material**

**Section 1: Extraction of symmetric and antisymmetric PHE**

To extract different contributions, the data acquired as a function of angle is taken at both positive and negative fields. Fig. S1(a) shows the raw data for the longitudinal resistance at 9 T, - 9 T and 310 K. The +9 T and -9 T curves are averaged to obtain the magnetoresistance (MR). Fig. S1(b) shows the corresponding data across the Hall contacts. A clear discrepancy is seen in the two data sets at +9 T and -9 T. A symmetric and antisymmetric Hall effect part is obtained from these. The symmetric part mostly consists of the planar Hall effect (PHE) alongside a small contribution from $R_{xx}$, due to a longitudinal mismatch in the Hall contacts. Using the MR from Fig. S1(a) and the symmetric part from Fig. S1(b), the PHE is obtained. As the PHE is also symmetric in magnetic field as the MR, it is not easy to separate the two. Using the rule that the PHE is antisymmetric as a function of angle, i.e., $\rho_{xy}^s(\theta) = -\rho_{xy}^s(\pi - \theta)$, we subtract a fraction of $R_{xx}$ (around 0.08 times) from $R_{xy}$ to obtain the PHE. However as shown in the main text, at low temperature this relationship is not strictly valid. For that, we have tried to ensure that the maximum and the minimum values in $R_{xy}$ are equal in magnitude but opposite in sign.

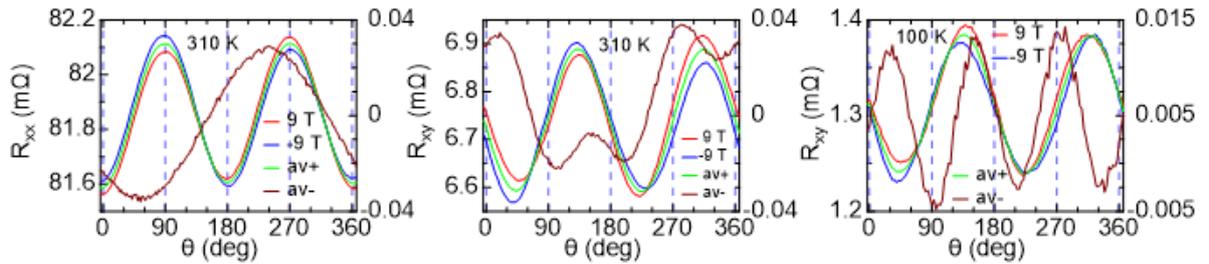

Fig S1. (a) Longitudinal and (b) transverse resistance measured at both positive and negative fields at 9 T and 310 K along with the average and antisymmteric terms. Y-axis for the antisymmteric term is on the right hand side. (c) Similar data as in (b) for 100 K and 9 T.

For the antisymmetric part of the PHE as in Fig. S2(a), we assume an out of plane contribution due to the sample plane being tilted relative to the plane of the rotation of the magnetic field and an in-plane term, representing a true antisymmetric PHE. The out of plane contribution should be sinusoidal at high magnetic field as the out-of-plane Hall effect is mostly due to the anomalous Hall effect (AHE). For the in-plane PHE we find that a triangular curve or a $sin3\theta$ function provides the best fitting. As noted in the main text, both functions provide qualitatively similar output. The value of $\rho_{xy}^a$ obtained from a triangular function fitting is 1.2 times from the corresponding value obtained using a $sin3\theta$ function fitting. The data in Fig. S2(a) is fitted to a sum of a triangle curve and a sine curve and shows a good fitting. The following function is used for fitting

$$\rho_{xy}^{a,total} = k + \rho_{AHE}\sin(\theta - \theta_A) + \rho_{xy}^a$$

$$\rho_{xy}^a = \rho_{xy}^a(0)\left(\left|mod\left[\frac{(\theta-\theta_o)}{30}, 4\right] - 2\right| - 1\right)$$

, where $k$ is an offset, $\rho_{AHE}$ is the AHE term, $\rho_{xy}^a$ is the triangular contribution, and $\rho_{xy}^a(0)$ is the amplitude of the triangular contribution. Alternatively, fitting can be done with

$$\rho_{xy}^{a,total} = k + \rho_{AHE}\sin(\theta - \theta_A) + \rho_{xy}^a(0)\sin3(\theta - \theta_o).$$

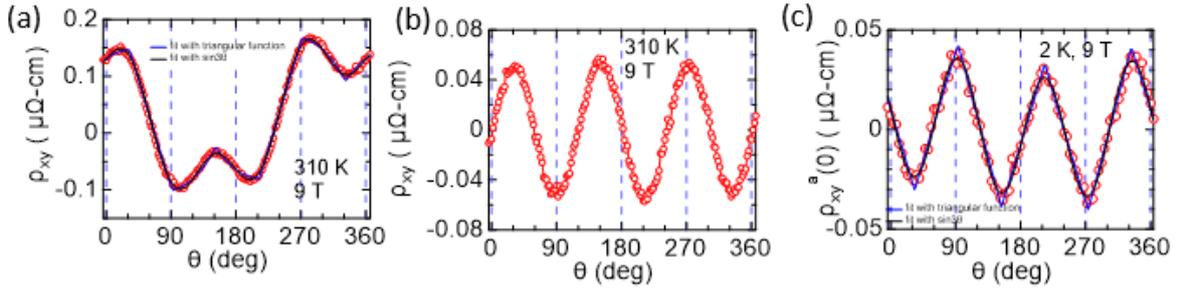

Fig S2. (a) The antisymmetric component of $\rho_{xy}$ at 310 K and 9 T. Fitting is with a triangle function and a sin3θ function. (b) Data after removing the sinusoidal contribution. (c) Data with fitting at 2 K and 9 T.

Fig. S2(b) shows the data after removing a constant offset and the sinusoidal contribution, clearly emphasizing the triangular nature of the antisymmetric PHE. At low temperatures such as 100 K, the triangular nature is seen in the data before fitting as seen in Fig. S1(c) due to the fact that the AHE is small at temperatures below the spin reorientation transition around 120 K. Fig. S2(c) shows the data at 2 K and 9 T with the fittings with the two formulas.

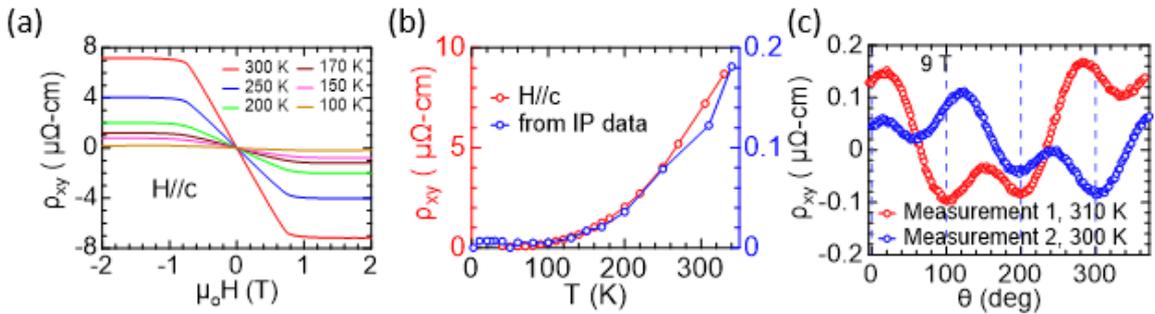

Fig S3. (a) The anomalous Hall effect at selected temperatures as a function of magnetic field. (b) Comparison between the anomalous Hall effect from (a) with the amplitude of the sinusoidal contribution in Fig. S2(a). (c) Comparison between two different measurements on the same sample under similar conditions. Sample was removed and remounted with new contacts for this.

To demonstrate that the sinusoidal contribution is indeed due to an AHE, we compared it with the Hall effect measured for $H \parallel \mathbf{c}$. Fig. S3(a) shows the Hall effect measured as a function of magnetic field at several temperatures. As demonstrated in Fig. S3(b), the temperature dependence of the sinusoidal term shows an excellent agreement with that of the AHE. Comparing the data between two different measurements on the same sample provides an additional confirmation. We conducted PHE measurements twice on the same sample, but with the sample being remounted with a set of new contacts during the second measurements. As seen in Fig. S3(c), the measured antisymmetric Hall resistance looks entirely different because the sample plane tilt with respect to the plane of the magnetic field rotation changes when we remount the sample, leading to a different amount of out-of-plane field and AHE contribution. Based on the sinusoidal component of the Hall resistance, the sample tilt was 1.04° and 0.59° for measurement 1 and 2, respectively. After subtracting the sinusoidal contribution due to the out-of-plane Hall effect, the in-plane term was the same for both measurements.

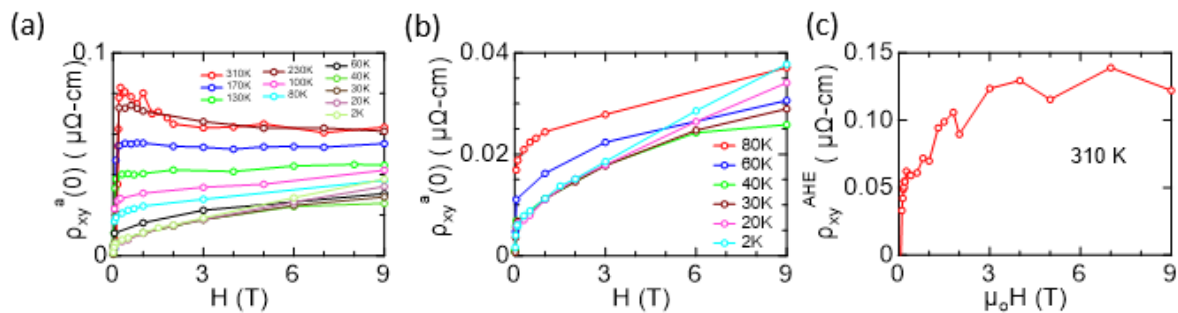

Fig S4. Antisymmetric planar Hall effect at several temperatures (a) from 310 to 2 K, (b) from 80 to 2 K. (c) Change in the anomalous Hall term with the magnetic field at 310 K.

In Fig. S4, we present some more data on the antisymmetric PHE vs $H$ at different temperatures. Especially Fig. S4(b) shows, how the antisymmetric PHE term develops with the magnetic field at lower temperature. Apart from a spontaneous jump at zero field, the antisymmetric PHE is linear in field. Fig. S4(c) is the change in AHE term with field at 310 K.

**Section 2: Additional PHE data**

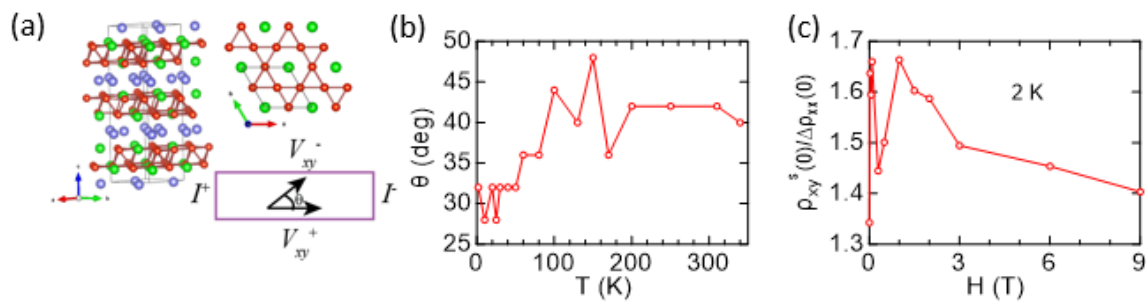

Fig S5. (a) Crystal structure of Fe$_3$Sn$_2$ and the schematic diagram of current and Hall contacts and the direction in which the field vector is rotated in the sample plane. (b) Change in the minima position in symmetric PHE at 9 T. (c) Change in the PHE to AMR ratio at 2 K with field.

Fig. S5(a) shows the measurement configuration with the magnetic field rotated in the plane of the sample along with the crystal structure of Fe$_3$Sn$_2$. Fig. S5(b) shows the change in the minima position in the symmetric PHE at 9 T as a function of temperature. This corresponds to the data in Fig. 2(a) in the main text. Fig. S5(c) shows the change in the PHE to AMR ratio at 2 K with field.

**Section 3: Dependence of various quantities on the magnetization**

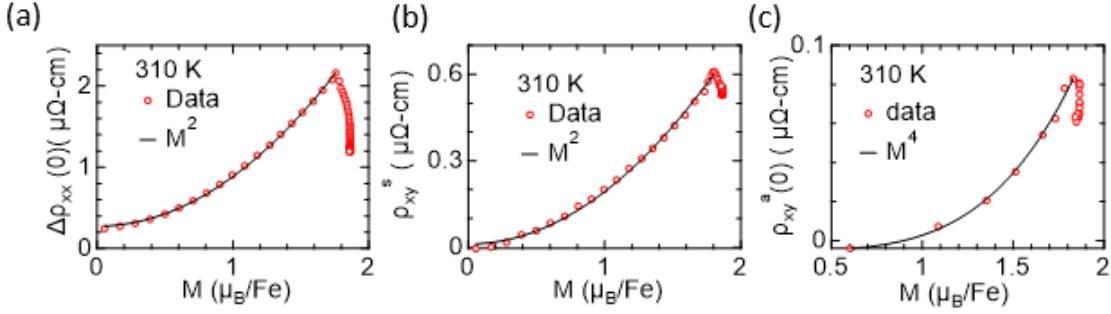

Fig S6. The dependence on the magnetization of (a) longitudinal AMR. (b) symmetric PHE at 9 T (c) antisymmetric PHE at 310 K.

Fig. S6 shows the dependence of various quantities on the magnetization as seen at 310 K. Although the magnetization data was recorded at 300 K for $H \parallel \mathbf{a}$, we expect the magnetization at 310 K to be the same as at 300 K.

**Section 4: Data scaled with the zero field resistivity**

Fig. S7(a)-(b) shows various quantities such as longitudinal AMR, symmetric PHE, and antisymmetric PHE measured at different temperatures normalized by the zero field resistivity $\rho_0$ and plotted as a function of $1/\rho_0$, or equivalently $\sigma_0$. They all exhibit a monotonic increase as $\sigma_0$ increases when the system is cooled down due to a larger mean free time. Fig. S7(c) shows the antisymmetric PHE at select temperatures normalized by $\rho_0$ as a function of magnetic field normalized by $\rho_0$ or $\frac{H}{\rho_0} \propto \omega_c \tau$. Although the data for high $\frac{H}{\rho_0}$ values do not scale well, with the data at different temperatures peeling off from each other, the jump at low $\frac{H}{\rho_0}$ appears to scale relatively well for $80\ K \leq T \leq 310\ K$ but forming a separate curve for $T \leq 40\ K$. The separation of the two sets of curves at $T = 40\ K$ could be related to the electronic transition detected at 40 K in $\rho_{xy}^a$ measured at 9 T.

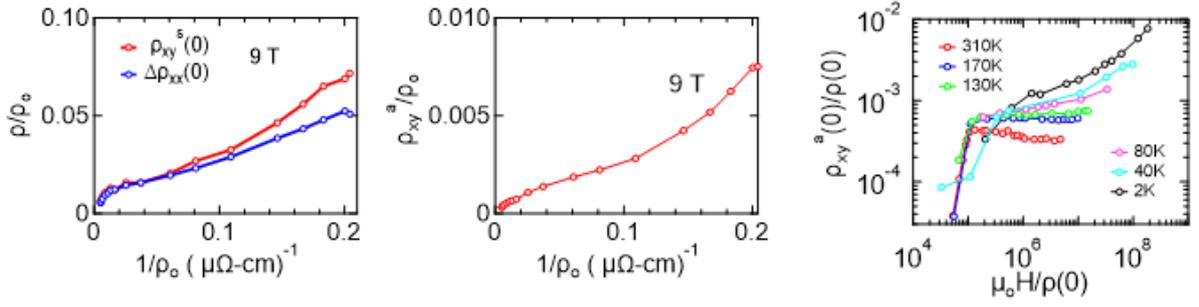

Fig S7. (a) Longitudinal AMR and symmetric PHE and (b) antisymmetric PHE at 9 T normalized by the zero field resistivity $\rho_0$ plotted as a function of $1/\rho_0$. (c) Antisymmetric PHE at select temperatures normalized by $\rho_0$ as a function of magnetic field normalized by $\rho_0$.

**Section 5: Data on an additional sample measured at room temperature**

To check the reproducibility of our finding, especially the antisymmetric PHE, we measured one more sample. In Fig. S8 and S9, we show the symmetric and antisymmetric PHE data obtained on this sample at 300 K. The symmetric PHE data shows a similar pattern as that presented in the main text, although the magnitude is larger. The antisymmetric PHE in Fig. S9(b) displays a 3-fold triangle curve with the same phase and has nearly the same magnitude as that presented in the main text. The tilt of the sample was higher in this case, which caused a very large contribution from the out-of-plane AHE, masking the in-plane component and introducing a large background noise to the antisymmetric PHE. By comparing this data with the out-of-plane AHE data, we calculated the tilt for this sample to be around 8°.

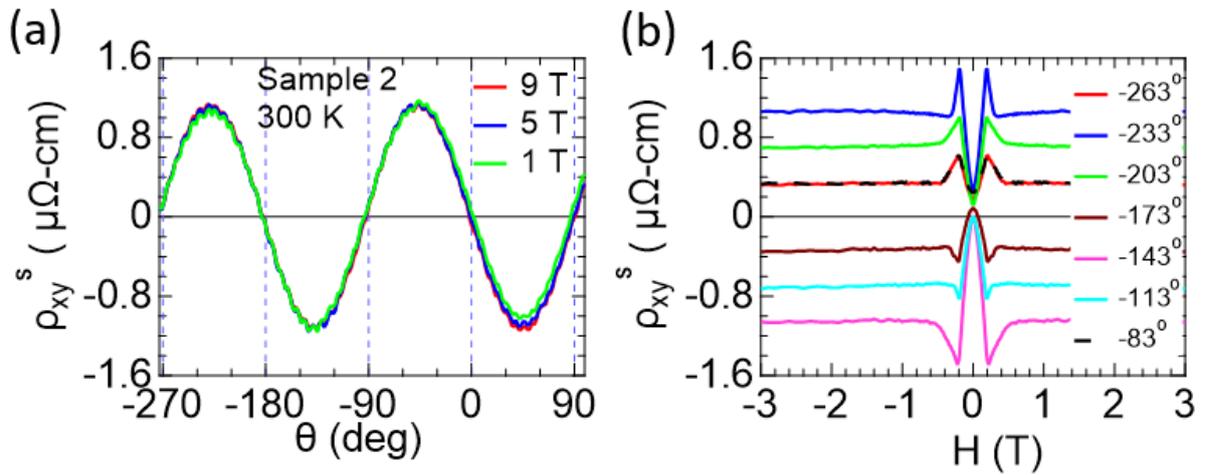

Fig S8. (a) Planar Hall effect as a function of angle at selected magnetic field for sample 2. (b) Planar Hall effect as a function of field at selected angles.

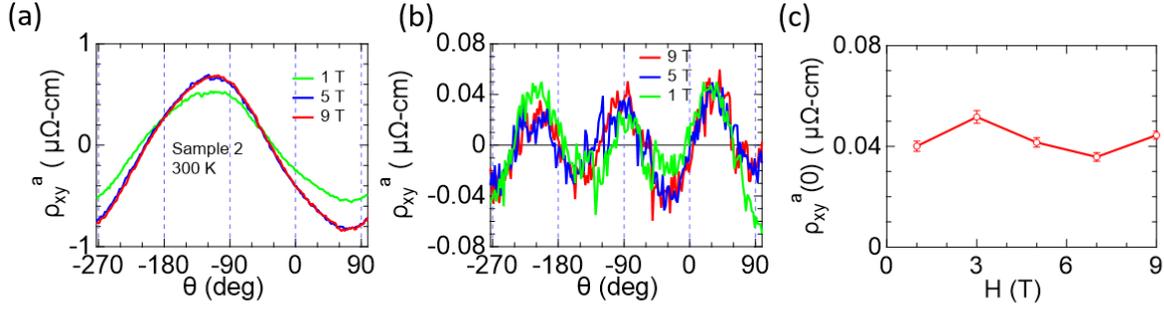

Fig S9. (a) The antisymmetric term at selected fields. (b) Antisymmetric PHE obtained after removing offset and sinusoidal term. (c) Amplitude of antisymmetric PHE as a function of magnetic field.

**Section 6: Data on sample 1 taken during measurement 2**

In order to ensure the reproducibility of our antisymmetric PHE, we measured sample 1 twice by mounting the sample second time with new contacts. The antisymmetric component of the Hall resistance had a different amount of sinusoidal contribution since the amount of tilt was different, 1.04° for the data presented in the main text compared to 0.59° for the original mount. As a result of the low amount of tilt in the original mount, the 3-fold symmetry is visible in Fig. S10 (a)-(b) even before subtracting the sinusoidal contribution due to the out-of-plane AHE. $\rho_{xy}^a$ at 300 K at 3 and 9 T as shown in Fig. S10(a), exhibits a 3-fold symmetry with maxima at 30°, 150°, and 270°, superimposed over a sinusoidal modulation. A maximum value of $\rho_{xy}^a$ of about 0.1 µΩ-cm is observed at 30°, with the corresponding minima at 210°. On decreasing the temperature, the maxima at 30° and minima at 210° decrease in magnitude while the other peaks and valleys remain largely unchanged as seen in Fig. S10(b). At around 150 K, all three peaks and three valleys become equal value, respectively. 150 K is also the temperature where the AHE for an out-of-plane magnetic field becomes very small. This is consistent with the sinusoidal modulation with $2\pi$ periodicity seen at higher temperatures originating from the AHE from an out-of-plane magnetic field component with the corresponding maxima and minima at 30° and 210°, respectively. Below 100 K, the behavior appears to be weakly dependent on temperature. To see it more clearly, we plot in Fig. S10(c) the temperature dependence of the 3-fold triangle peak amplitude obtained by fitting all the curves. The PHE decreases with decreasing temperature after showing a peak at 250 K. At 40 K, it shows a minimum before increasing with decreasing temperature, which suggests an electronic transition. Fig. S10(d) shows $\rho_{xy}^a$ as a function of azimuthal angle at 2 K for various magnetic fields showing an increase as the field is increased.

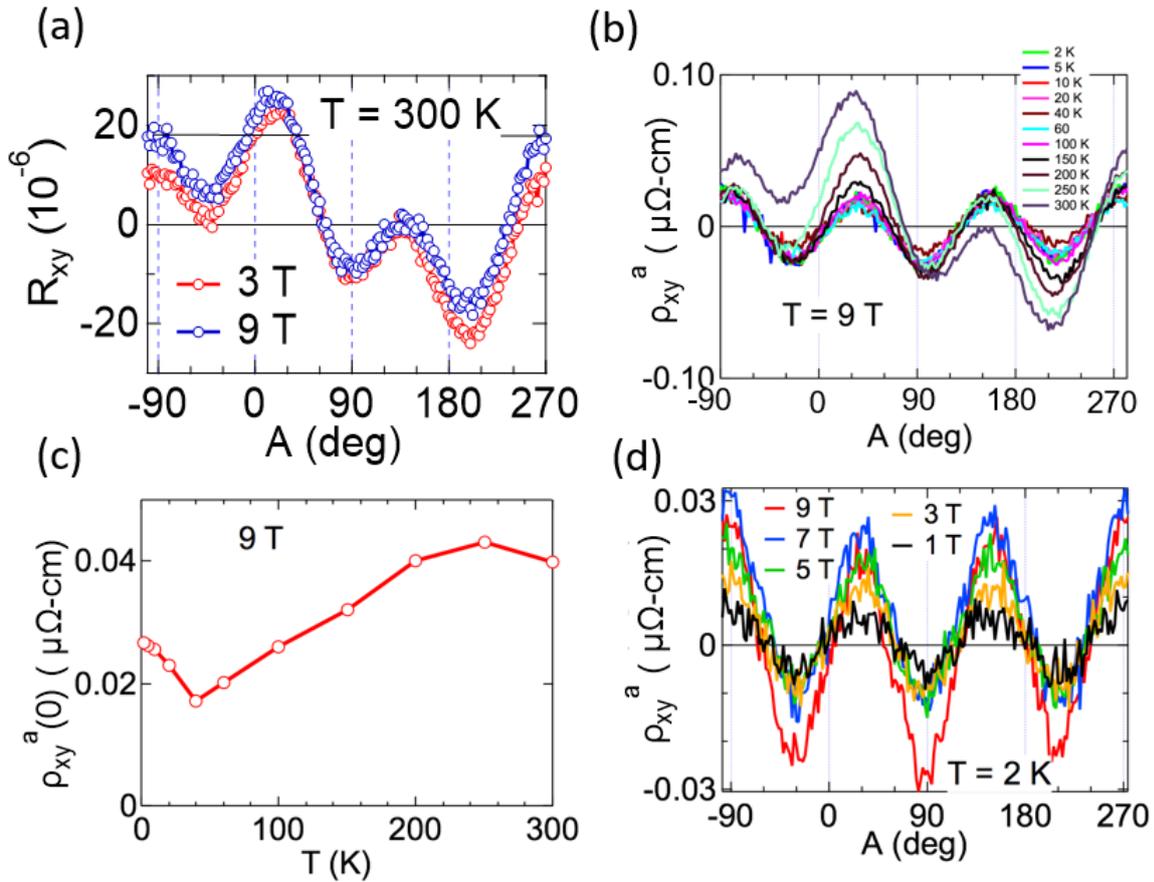

Fig S10. Data taken during measurement 2 on sample 1. Antisymmetric planar Hall effect (a) at 300 K. (b) at 9 T at various temperatures, (c) antisymmetric planar Hall effect amplitude as a function of temperature, showing an increase at low temperature. (d) Antisymmetric planar Hall effect at 2 K and various magnetic field values.

7